\documentclass[a4paper,11pt]{article}
\usepackage{jheppub}
\usepackage{amsmath,amssymb} % for details on the use of the package, please
\usepackage{graphicx}
\usepackage{xcolor}

\usepackage{tikz}
\usetikzlibrary{matrix,arrows,shapes,decorations.fractals,decorations.pathmorphing,shadows,patterns,decorations.markings}
\definecolor{yellow}{cmyk}{0,0,1,0}
\definecolor{magenta2}{cmyk}{0.25,0.8,0,0.1}
\definecolor{cyan2}{cmyk}{0.9,0.02,0,0.1}
\definecolor{green2}{cmyk}{1,0,1,0.15}
\definecolor{oran}{cmyk}{0,0.2,0.57,0}

\usepackage{subcaption}

\title{\boldmath Spin chain integrability in non-supersymmetric Wilson loops}

\author[a]{Diego Correa,}
\author[b]{Matias Leoni,}%\note{Corresponding author.}}
\author[b]{Solange Luque}

\affiliation[a]{Instituto de F\'isica La Plata, CONICET, Universidad Nacional de La Plata C.C. 67, 1900 La Plata, Argentina}
\affiliation[b]{Departamento de F\'{\i}sica, Universidad de Buenos Aires \& IFIBA - CONICET Ciudad Universitaria, pabell\'on 1 (1428) Buenos Aires, Argentina}
\emailAdd{correa@fisica.unlp.edu.ar}
\emailAdd{leoni@df.uba.ar}
\emailAdd{sluque@df.uba.ar}
\abstract{We study the 1-loop dilatation operator for insertions of composite operators in
a generalized Wilson loop in ${\cal N}=4$ super Yang-Mills, which interpolates
between the supersymmetric Wilson-Maldacena loop and the ordinary Wilson loop with
 no scalar coupling. For $SO(6)$ scalar insertions, we show that the 1-loop
dilatation operator is integrable for the endpoints of the interpolation,
{\it i.e.} either for the Wilson-Maldacena or the ordinary Wilson loop. Moreover,
we also show that integrability persists for $SU(2|3)$ insertions in the ordinary
Wilson loop, even when the term making the spin chain length dynamical is included.}

\begin{document}
\maketitle
\flushbottom

\section{Introduction}

One of the most studied operators in $\mathcal{N}=4$ Super-Yang-Mills is the so called Wilson-Maldacena loop (WML) operator \cite{Maldacena:1998im,Rey:1998ik}
\begin{equation}
W(C)=\frac{1}{N}\, \mathrm{Tr}\mathcal{P}\,\exp
\oint\limits_{C} d\tau \left(i A_{\mu}(x)\dot{x}^{\mu}+|\dot{x}|n_{I}\Phi^{I}(x)\right)
\end{equation}
with $n_I$ a unit six-vector. It is a gauge invariant non-local operator which preserves half of the supersymmetries locally. Its expectation value on a straight line and for a constant $n_I$ is trivial and it is exactly known for a circle contour to all orders in the coupling $g_{Y\!M}$ and the number of colors $N$ by means of localization \cite{Erickson:2000af,Drukker:2000rr,Pestun:2007rz}. Of the many interesting results regarding this operator, one could mention for example those in \cite{Correa:2012at} where they relate the logarithmic derivative of the expectation value of this Wilson loop, known exactly, to the function controlling the small angle cusp anomalous dimension. Moreover in \cite{Correa:2012hh,Drukker:2012de}, by using a spin chain construction a TBA equation was given which determines the full cusp anomalous dimension controlling the divergences on a cusp of this Wilson loop. There exist many other interesting results related to this Wilson loop (see {\it e.g.} \cite{Gromov:2012eu,Gromov:2015dfa,Gromov:2013qga,Giombi:2012ep,Kim:2017phs}).

In part for not being supersymmetric, much less attention was paid to the ordinary  Wilson loop (WL) which contains no coupling to the adjoint scalars of the theory. In fact, most of the exact results described in the last paragraph are unknown for the ordinary operator. But as argued in \cite{Polchinski:2011im}, the ordinary Wilson loop is an interesting object in its own right even in the context of a supersymmetric field theory. There, it was shown that since the WML and the ordinary WL map through AdS/CFT to string worldsheets with Dirichlet and Neumann boundary conditions for world-sheet fields respectively, a renormalization group flow should exist between both operators with the WL in the ultraviolet and the WML in the infrared. The coupling $\zeta$ is introduced in the modified operator
\begin{equation}\label{ZetaWL}
W^{(\zeta)}(C)=\frac{1}{N}\, \mathrm{Tr}\mathcal{P}\,\exp
\oint\limits_{C} d\tau \left(i A_{\mu}(x)\dot{x}^{\mu}+\zeta\,|\dot{x}|n_{I}\Phi^{I}(x)\right)
\end{equation}
such than $W^{(0)}(C)$ is the ordinary WL while $W^{(1)}(C)$ is the WML operator. Taking a circular contour and fixed $n_{I}=\delta_I^4$ one can think of the expectation value of this operator as the partition function of some one-dimensional defect quantum field theory on $S^1$ perturbed from its conformal points by a weakly relevant operator (the coupling to the scalar $\Phi^4$) which drives the renormalization flow. At the perturbative level, the $\zeta$ coupling has to be renormalized and the following renormalization flow is found \cite{Polchinski:2011im}
\begin{equation}\label{BetaFunction}
\beta_{\zeta}(\lambda,\zeta)=\mu\frac{\partial\zeta}{\partial\mu}=-\frac{\lambda}{16\pi^2}\zeta(1-\zeta^2)+\mathcal{O}(\lambda^2), \qquad \lambda=g_{Y\!M}^2 N
\end{equation}
which has fixed points at $\zeta=0$ and $\zeta=\pm 1$ . From (\ref{BetaFunction}) we see that the ordinary WL is a UV fixed point while the WML is an IR fixed point.

The authors of \cite{Beccaria:2017rbe} took a further step and made the explicit computation of the deformed Wilson loop (\ref{ZetaWL}) for the circular contour up to second order in perturbation theory using dimensional regularization. After rewriting the result in terms of the renormalized coupling $\zeta$ through (\ref{BetaFunction}) they obtained
\begin{equation}\label{WilsonZeta}
\langle W^{(\zeta)}\rangle=
1+\frac{1}{8}\lambda+\left(\frac{1}{192}+\frac{1}{128\pi^2}(1-\zeta^2)^2\right)\lambda^2+\mathcal{O}(\lambda^3)
\end{equation}
We see that the result is independent of $\zeta$ at leading order and the first non-trivial result appears at second order. It matches the known expectation value for $\zeta=1$, $\langle W^{(1)}\rangle=2\lambda^{-\tfrac{1}{2}}I_1(\sqrt{\lambda})=1+\tfrac{\lambda}{8}+\tfrac{\lambda^2}{192}+\mathcal{O}(\lambda^3)$ and breaks uniform transcendentality at $\zeta=0$, $\langle W^{(0)}\rangle=1+\tfrac{\lambda}{8}+\left(\tfrac{1}{192}+\tfrac{1}{128\pi^2}\right)\lambda^2+\mathcal{O}(\lambda^3)$.

Other results can be related to (\ref{WilsonZeta}). If we insist on the interpretation of an underlying one dimensional (defect) quantum field theory on $S^1$ or the straight line, as we will consider in this paper, interesting objects are correlators of gauge covariant operators defined by
\begin{equation}\label{CorrelationFunctions}
\langle\!\langle O(\tau_n)\dots O(\tau_1)\rangle\!\rangle^{(\zeta)}=
\frac{\langle \mathrm{Tr}(\mathcal{P}O(\tau_n)\dots O(\tau_1)e^{\int d\tau (iA.\dot{x}+\zeta|\dot{x}|\Phi^4)})\rangle}{\langle \mathrm{Tr}\mathcal{P}e^{\int d\tau (iA.\dot{x}+\zeta|\dot{x}|\Phi^4)}\rangle}
\end{equation}
where a Wilson exponential is to be understood between the different operators. For the fixed points $\zeta=0,\pm 1$ these would be correlators in a one dimensional defect conformal field theory (CFT) and are particularly interesting since they could be used to understand another example of an AdS$_2$/CFT$_1$ duality like the ones discussed in \cite{Cooke:2017qgm,Giombi:2017cqn}. From conformal covariance on $\mathbb{R}$ one expects for an operator with dimension $\Delta$ that the two and three points correlators have a functional dependence on the insertion points given by (only for $\zeta=0,\pm 1$)
\begin{align}\label{twothreepoints}
&\langle\!\langle O_{\Delta}(\tau_2) O_{\Delta}(\tau_1)\rangle\!\rangle_{\mathrm{line}}=\frac{B(\lambda)}{(\tau_2-\tau_1)^{2\Delta}}
\nonumber\\
&\langle\!\langle O_{\Delta}(\tau_3) O_{\Delta}(\tau_2) O_{\Delta}(\tau_1)\rangle\!\rangle_{\mathrm{line}}=\frac{C(\lambda)}{(\tau_3-\tau_1)^{2\Delta}(\tau_2-\tau_1)^{2\Delta}(\tau_3-\tau_2)^{2\Delta}}
\end{align}
For the WML case and the operator $O=\Phi^4$ the two point function was explicitly computed in \cite{Alday:2007he} obtaining for the dimension $\Delta_4=1+\tfrac{\lambda}{4\pi^2}+\mathcal{O}(\lambda^2)$, while the operators $\Phi^I$ with $I\neq 4$ are protected. In fact the all loop spectral problem of operators in the WML was { in principle solved in \cite{Correa:2012hh,Drukker:2012de} using integrability}. For the WL case on the other hand, all single scalar operators have the same dimension given by  $\Delta=1-\tfrac{\lambda}{8\pi^2}+\mathcal{O}(\lambda^2)$. These results can also be checked by expanding the definition of the Wilson loop (\ref{ZetaWL}) and comparing it with its expansion (\ref{WilsonZeta}) as done in \cite{Beccaria:2017rbe}. There they were also able to use the Wilson loop expectation value to determine the leading order of structure constants appearing in the three point function (\ref{twothreepoints}) for the $\Phi^4$ operator. For recent works in non-supersymmetric Wilson loops and underlying defect CFTs see \cite{Beccaria:2018ocq,Hoyos:2018jky,Bianchi:2018zpb,Giombi:2018qox}.

Motivated by these results we tackle the spectral problem of  composite operators living in the Wilson loop which depends on the parameter $\zeta$. In the first place, we will derive and study the corresponding open spin chain Hamiltonian that controls the mixing of the set of all scalar composite operators of length $L$
\begin{equation}
\Phi^{I_1}\Phi^{I_2}....\Phi^{I_L}(\tau)
\end{equation}
inserted in a straight Wilson line and characterized by a ``word'' of flavours $I_l=1,...,6$. This $SO(6)$ sector includes the $\Phi^4$ scalar which interacts with the Wilson loop scalar insertion already at one loop. This constitutes a twofold generalization of the problem analyzed in \cite{Drukker:2006xg}: (i) the Wilson loop specifying the one-dimensional defect is the $\zeta$-deformed one and (ii) the composite operators are taken in the larger sector of $SO(6)$. This is a sector which is big enough so that it includes non-trivial constraints for integrability but excludes other possible operators worth studying.

The lack of supersymmetry  in the case of  Wilson loops with $\zeta\neq\pm 1$ should not be considered a priori as an impediment to the integrability of the system. Although the  fermionic symmetries of the superconformal $PSU(2,2|4)$ are known to play a central role in the determination of integrable bulk scattering and reflection matrices, there are examples of ${\cal N}=4$ super Yang-Mills deformations that, while breaking supersymmetry completely, do not spoil integrability \cite{Frolov:2005dj}.

The work is organized as follows. In the next section we use the spin chain analogy and we establish the one loop Hamiltonian which is the mixing operator of the $SO(6)$ closed sector in addition to some boundary terms. For this, we recycle known spin chain results and we compute the novel diagrams which contribute to the boundary Hamiltonian. In section 3 we obtain the energies (anomalous dimensions), scattering matrix and the reflection matrix --which will be dependent on the parameter $\zeta$ of the scalar insertions-- for this spin chain by using the coordinate Bethe ansatz for one and two magnon excitations. After evaluating the boundary Yang-Baxter equation we find that it is only fulfilled as long as
\begin{equation}
\zeta=0,\qquad \zeta=\pm 1
\end{equation}
Thus, we see that system is one loop integrable only for the fixed points of the renormalization flow at one loop, that is, for the WML and for the ordinary WL where a one dimensional CFT underlying the correlators exists. Then, and looking for a more stringent test of integrability, we consider $SU(2|3)$ insertions in an ordinary $\zeta =0$  WL and verify that integrability holds in this other sector. Finally, we conclude in section 4 with a discussion of our results. Details of some contour integrals have been relegated to the appendix.

\section{One loop bulk and boundary dilatation operator}

We start by considering the set of all scalar operators of the form $O^{I_1...I_L}=\Phi^{I_1}...\Phi^{I_L}$ inserted in the straight Wilson line parametrized by $x^{\mu}=(\tau,0,0,0)$. With a single scalar propagator given by $\tfrac{g_{Y\!M}^2}{4\pi^2|x-y|^2}$ we obtain for the tree level planar two point function
\begin{equation}\label{TreeLevel}
\langle\!\langle
O^{I_1,I_2,...,I_L}(\tau)\bar{O}_{J_L,J_{L\!-\!1},...,J_1}(0)\rangle\!\rangle_{\mathrm{tree}}=
\left(\frac{\lambda}{8\pi^2}\right)^L
\frac{1}{\tau^{2L}}
\ \delta^{I_1}_{J_1}...\delta^{I_L}_{J_L}=\mathcal{T}_L(\tau)\ \delta^{I_1}_{J_1}...\delta^{I_L}_{J_L}
\end{equation}
which, due to the planarity of the scalar contractions, is proportional to the `identity' $\delta^{I_1}_{J_1}\delta^{I_2}_{J_2}...\delta^{I_L}_{J_L}$ in the flavour space.

At loop level, quantum corrections will appear in the form of divergencies which we choose to regularize with dimensional regularization. This amounts to turning integrals $\int d^4x\to\mu^{-2\epsilon}\int d^D x$ with $D=4-2\epsilon$ and using regularized propagators such as the real scalar and gluon one (in the Feynman gauge and omitting color/flavour indexes)
\begin{equation}
\langle\Phi^I(x)\Phi^J(y)\rangle=\frac{\Gamma(1-\epsilon)g_{Y\!M}^2}{4\pi^{2-\epsilon}}
\frac{\mu^{2\epsilon}\delta^{IJ}}{((x-y)^2)^{1-\epsilon}}\quad
\langle A_\mu(x) A_\nu(y)\rangle=\frac{\Gamma(1-\epsilon)g_{Y\!M}^2}{4\pi^{2-\epsilon}}
\frac{\mu^{2\epsilon} \eta_{\mu\nu}}{((x-y)^2)^{1-\epsilon}}
\end{equation}
with $\mu$ being 't Hooft mass. We will absorb divergencies in a renormalization matrix $\mathcal{Z}$ which mixes the operators $O^{a}_{\mathrm{bare}}=\mathcal{Z}^a_b O^{b}_{\mathrm{ren}}$, has a perturbative expansion $\mathcal{Z}=1+\lambda\mathcal{Z}^{(1)}+\mathcal{O}(\lambda^2)$ and can be computed as minus the sum of the poles in the $\epsilon$ expansion of Feynman diagram contributions modulo the tree level contribution. The dilatation operator, which we aim to obtain and diagonalize, is given by
\begin{equation}
\mathcal{D}=\mu\frac{d}{d\mu}\log\mathcal{Z}
\end{equation}
and since $\mu$ will always appear in the form $\lambda\mu^{2\epsilon}$ and at leading order $\log\mathcal{Z}=\mathcal{Z}^{(1)}\lambda+\mathcal{O}(\lambda^2)$ we may write
\begin{equation}
\mathcal{D}=\lim_{\epsilon\to 0}
\left(2\epsilon
\lambda\frac{d}{d\lambda}\log\mathcal{Z}
\right)=2\epsilon \mathcal{Z}^{(1)}\lambda+\mathcal{O}(\lambda^2)
\end{equation}
thus, effectively, the one loop dilatation $\mathcal{D}^{(1)}$ operator is given by two times the simple pole of the $\mathcal{Z}^{(1)}$ renormalization constant.

In order to diagonalize the dilatation operator, we think of it as a Hamiltonian acting in a space which is the product of $L$ Hilbert spaces of the $SO(6)$ vector representation. The dilatation operator will decompose in three types of contributions
\begin{equation}
\mathcal{D}^{(1)}= H^{(1)}(\zeta) = E_0 + H_{\mathrm{bulk}} +H_{\mathrm{bdry}}
\end{equation}

\begin{figure}[!ht]
    \centering
\begin{subfigure}[b]{0.15\textwidth}
\begin{tikzpicture}[scale=0.8]
\draw  (0,-2) -- (0,2);
\draw[thick] (0,-1) arc (-90:90: 1 and 1) ;
\draw[thick] (0,-1) arc (-90:90: 0.92 and 1) ;
\draw[thick] (0,-1) arc (-90:90: 0.84 and 1) ;
\draw[dotted] (0,-1) arc (-90:90: 0.75 and 1) ;
\draw[thick] (0,-1) arc (-90:90: 0.66 and 1) ;
\draw[thick] (0,-1) arc (-90:90: 0.58 and 1) ;
\draw[thick] (0,-1) arc (-90:90: 0.5 and 1) ;
\draw[thick,dashed] (0,-1.5) arc (-90:90: 1.5 and 1.5) ;
\fill[white] (0,1.5) circle (1.5pt);
\draw (0-0.033,1.5-0.033)--(0+0.033,1.5+0.033);
\draw (0,1.5) circle (1.5pt);
\fill[white] (0,-1.5) circle (1.5pt);
\draw (0-0.033,-1.5-0.033)--(0+0.033,-1.5+0.033);
\draw (0,-1.5) circle (1.5pt);
\draw (-0.05,-1) node[left] {$0$};
\draw (-0.05,1) node[left] {$\tau$};
\draw (-0.05,-1.5) node[left] {$\tau_1$};
\draw (-0.05,1.5) node[left] {$\tau_2$};
\end{tikzpicture}
        \caption{}
        \label{fig:Ha}
    \end{subfigure}\!\!\!
        \begin{subfigure}[b]{0.15\textwidth}
\begin{tikzpicture}[scale=0.8]
\draw  (0,-2) -- (0,2);
\draw[thick] (0,-1) arc (-90:90: 1 and 1) ;
\draw[thick] (0,-1) arc (-90:90: 0.92 and 1) ;
\draw[thick] (0,-1) arc (-90:90: 0.84 and 1) ;
\draw[dotted] (0,-1) arc (-90:90: 0.75 and 1) ;
\draw[thick] (0,-1) arc (-90:90: 0.66 and 1) ;
\draw[thick] (0,-1) arc (-90:90: 0.58 and 1) ;
\draw[thick] (0,-1) arc (-90:90: 0.5 and 1) ;
\draw[thick,dashed] (0,-0.5) arc (-90:90: 0.25 and 0.5) ;
\fill[white] (0,0.5) circle (1.5pt);
\draw (0-0.033,0.5-0.033)--(0+0.033,0.5+0.033);
\draw (0,0.5) circle (1.5pt);
\fill[white] (0,-0.5) circle (1.5pt);
\draw (0-0.033,-0.5-0.033)--(0+0.033,-0.5+0.033);
\draw (0,-0.5) circle (1.5pt);
\draw (-0.05,-1) node[left] {$0$};
\draw (-0.05,1) node[left] {$\tau$};
\draw (-0.05,-0.5) node[left] {$\tau_1$};
\draw (-0.05,0.5) node[left] {$\tau_2$};
\end{tikzpicture}
        \caption{}
        \label{fig:Hb}
    \end{subfigure}\!\!\!
        \begin{subfigure}[b]{0.15\textwidth}
\begin{tikzpicture}[scale=0.8]
\draw  (0,-2) -- (0,2);
\draw[thick] (0,-1) arc (-90:90: 1 and 1) ;
\draw[thick] (0,-1) arc (-90:90: 0.92 and 1) ;
\draw[thick] (0,-1) arc (-90:90: 0.84 and 1) ;
\draw[dotted] (0,-1) arc (-90:90: 0.75 and 1) ;
\draw[thick] (0,-1) arc (-90:90: 0.66 and 1) ;
\draw[thick] (0,-1) arc (-90:90: 0.58 and 1) ;
\draw[thick] (0,-1) arc (-90:90: 0.5 and 1) ;
\draw[thick] (0,1) arc (90:-90: 0.2 and 0.35) ;
\draw[thick] (0,-1) arc (90:-90: 0.2 and 0.35) ;
\fill[white] (0,0.3) circle (1.5pt);
\draw (0-0.033,0.3-0.033)--(0+0.033,0.3+0.033);
\draw (0,0.3) circle (1.5pt);
\fill[white] (0,-1.7) circle (1.5pt);
\draw (0-0.033,-1.7-0.033)--(0+0.033,-1.7+0.033);
\draw (0,-1.7) circle (1.5pt);
\draw (-0.05,-1) node[left] {$0$};
\draw (-0.05,1) node[left] {$\tau$};
\draw (-0.05,-1.7) node[left] {$\tau_1$};
\draw (-0.05,0.3) node[left] {$\tau_2$};
\end{tikzpicture}
        \caption{}
        \label{fig:Hc}
    \end{subfigure}\!\!\!
    \begin{subfigure}[b]{0.15\textwidth}
\begin{tikzpicture}[scale=0.8]
\draw  (0,-2) -- (0,2);
\draw[thick] (0,-1) arc (-90:90: 1 and 1) ;
\draw[thick] (0,-1) arc (-90:90: 0.92 and 1) ;
\draw[thick] (0,-1) arc (-90:90: 0.84 and 1) ;
\draw[dotted] (0,-1) arc (-90:90: 0.75 and 1) ;
\draw[thick] (0,-1) arc (-90:90: 0.66 and 1) ;
\draw[thick] (0,-1) arc (-90:90: 0.58 and 1) ;
\draw[thick] (0,-1) arc (-90:90: 0.5 and 1) ;
\draw[thick] (0,1.) arc (-90:90: 0.2 and 0.35) ;
\draw[thick] (0,-1) arc (90:-90: 0.2 and 0.35) ;
\fill[white] (0,1.7) circle (1.5pt);
\draw (0-0.033,1.7-0.033)--(0+0.033,1.7+0.033);
\draw (0,1.7) circle (1.5pt);
\fill[white] (0,-1.7) circle (1.5pt);
\draw (0-0.033,-1.7-0.033)--(0+0.033,-1.7+0.033);
\draw (0,-1.7) circle (1.5pt);
\draw (-0.05,-1) node[left] {$0$};
\draw (-0.05,1) node[left] {$\tau$};
\draw (-0.05,-1.7) node[left] {$\tau_1$};
\draw (-0.05,1.7) node[left] {$\tau_2$};
\end{tikzpicture}
        \caption{}
        \label{fig:Hd}
    \end{subfigure}\!\!\!
        \begin{subfigure}[b]{0.15\textwidth}
\begin{tikzpicture}[scale=0.8]
\draw  (0,-2) -- (0,2);
\draw[thick] (0,-1) arc (-90:90: 1 and 1) ;
\draw[thick] (0,-1) arc (-90:90: 0.92 and 1) ;
\draw[thick] (0,-1) arc (-90:90: 0.84 and 1) ;
\draw[dotted] (0,-1) arc (-90:90: 0.75 and 1) ;
\draw[thick] (0,-1) arc (-90:90: 0.66 and 1) ;
\draw[thick] (0,-1) arc (-90:90: 0.58 and 1) ;
\draw[thick] (0,-1) arc (-90:90: 0.5 and 1) ;
\draw[thick] (0,1) arc (90:-90: 0.2 and 0.35) ;
\draw[thick] (0,-1) arc (-90:90: 0.2 and 0.35) ;
\fill[white] (0,0.3) circle (1.5pt);
\draw (0-0.033,0.3-0.033)--(0+0.033,0.3+0.033);
\draw (0,0.3) circle (1.5pt);
\fill[white] (0,-0.3) circle (1.5pt);
\draw (0-0.033,-0.3-0.033)--(0+0.033,-0.3+0.033);
\draw (0,-0.3) circle (1.5pt);
\draw (-0.05,-1) node[left] {$0$};
\draw (-0.05,0.3) node[left] {$\tau_2$};
\draw (-0.05,-0.3) node[left] {$\tau_1$};
\draw (-0.05,1.) node[left] {$\tau$};
\end{tikzpicture}
        \caption{}
        \label{fig:He}
    \end{subfigure}\!\!\!
            \begin{subfigure}[b]{0.15\textwidth}
\begin{tikzpicture}[scale=0.8]
\draw  (0,-2) -- (0,2);
\draw[thick] (0,-1) arc (-90:90: 1 and 1) ;
\draw[thick] (0,-1) arc (-90:90: 0.92 and 1) ;
\draw[thick] (0,-1) arc (-90:90: 0.84 and 1) ;
\draw[dotted] (0,-1) arc (-90:90: 0.75 and 1) ;
\draw[thick] (0,-1) arc (-90:90: 0.66 and 1) ;
\draw[thick] (0,-1) arc (-90:90: 0.58 and 1) ;
\draw[thick] (0,-1) arc (-90:90: 0.5 and 1) ;
\draw[thick] (0,1.) arc (-90:90: 0.2 and 0.35) ;
\draw[thick] (0,-1) arc (-90:90: 0.2 and 0.35) ;
\fill[white] (0,1.7) circle (1.5pt);
\draw (0-0.033,1.7-0.033)--(0+0.033,1.7+0.033);
\draw (0,1.7) circle (1.5pt);
\fill[white] (0,-0.3) circle (1.5pt);
\draw (0-0.033,-0.3-0.033)--(0+0.033,-0.3+0.033);
\draw (0,-0.3) circle (1.5pt);
\draw (-0.05,-1) node[left] {$0$};
\draw (-0.05,1) node[left] {$\tau$};
\draw (-0.05,-0.3) node[left] {$\tau_1$};
\draw (-0.05,1.7) node[left] {$\tau_2$};
\end{tikzpicture}
        \caption{}
        \label{fig:Hf}
    \end{subfigure}

    \caption{Feynman diagrams contributing to $E_0$ and $H_{bdry}$. The thin vertical lines are the Wilson paths while the thick lines are scalar propagators. The dashed line is both a scalar and a gluon propagator. Operators are positioned at $0$ and $\tau$. Small circular vertices are drawn in the scalar/gluon insertions of the Wilson loop.}
\end{figure}
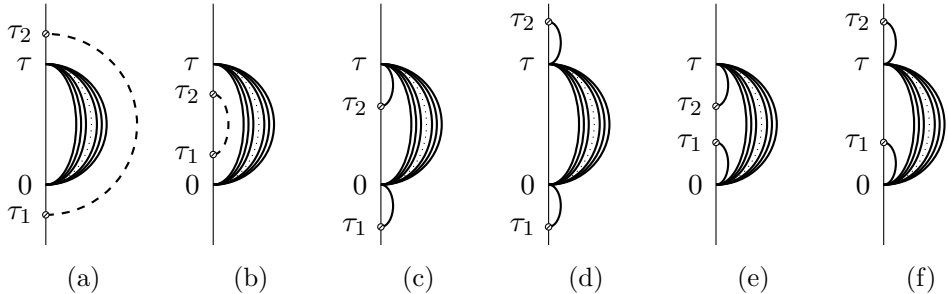

The first contribution $E_0$ comes from the diagrams shown in figures \ref{fig:Ha} and \ref{fig:Hb}. In those diagrams all the scalars from the operators are contracted as at tree level and we have propagators connecting the Wilson loop with itself. The reader might wonder whether these contributions should cancel with the expectation value normalization we used to define the correlator in (\ref{CorrelationFunctions}). As explained in \cite{Alday:2007he}, the planarity restriction forbids us to connect the upper part of the diagram with its lower part and therefore this cancelation is not produced. Moreover, while the exchange of a scalar/gluon within the Wilson loop when we go all along the line does not have a logarithmic divergence, this exchange does have this kind of divergence when we can not go all along due to the obstruction produced by the inserted operators. Its contribution to the two point function is
\begin{equation}
\frac{\lambda}{4\pi^2}
\mathcal{T}_L(\tau)
\delta^{I_1}_{J_1}...\delta^{I_L}_{J_L} \int\limits_0^\tau d\tau_2\int\limits_0^{\tau_2}d\tau_1
\frac{\zeta^2-1}{\left(\tau_2-\tau_1\right)^{2-2\epsilon}} =
- \frac{\lambda}{8\pi^2}(\zeta^2-1) \frac{1}{\epsilon} \mathcal{T}_L(\tau)
\delta^{I_1}_{J_1}...\delta^{I_L}_{J_L}+\mathcal{O}(\epsilon^0)
\end{equation}
where $\mathcal{T}_L(\tau)$ was implicitly defined in (\ref{TreeLevel}).
Thus, the contribution $E_0$ is just a constant and does not mix operators
\begin{equation}
E_0=\frac{\lambda}{8\pi^2}(\zeta^2-1)
\end{equation}
such that it vanishes for $\zeta=\pm 1$, {\it i.e.} for the WML. It should be noticed that this `vacuum energy' coming from diagrams which only involve the Wilson loop is not physical in the sense that it is gauge dependent. In particular, similar to what it was observed in \cite{Hoyos:2018jky}, in the Yennie gauge this contribution would have been $E_0=\frac{\lambda}{8\pi^2}\zeta^2$ thus vanishing for the ordinary WL instead.

The next set of contributions comes from the diagrams involving either self energy corrections of the scalar propagators or interactions between nearest neighbour scalar propagators. These corrections were already computed in \cite{Minahan:2002ve} in the context of $SO(6)$ single trace operators. The only difference in our case is that since our operators are not single trace (and thus our spin chain is not periodic), there are some extra diagrams since we do not have nearest neighbour interactions between the first and the last propagator. This was already observed in \cite{Drukker:2006xg} where it was shown that this excess is compensated by a diagram in which there is an exchange of a gluon between the outermost scalar propagators and the Wilson loop. These contributions constitute the `bulk Hamiltonian' of our open $SO(6)$ spin chain \cite{Minahan:2002ve} :
\begin{equation}
H_{\mathrm{bulk}}=\frac{\lambda}{8\pi^2}\sum\limits_{l=1}^{L-1}\left(1-\mathcal{P}_{l,l+1}+\frac{1}{2}\mathcal{K}_{l,l+1}\right)
\end{equation}
where $\mathcal{P}_{l,l+1}$ is the permutation operator which acts as
%on the flavour identity as
\begin{equation}
\left(\mathcal{P}_{l,l+1}\right)^{I_1 I_2 \cdots I_L}_{J_1 J_2 \cdots J_L}
%\ \delta^{I_1}_{J_1}\dots\delta^{I_l}_{J_l}\delta^{I_{l+1}}_{J_{l+1}}\dots\delta^{I_L}_{J_L}
=\delta^{I_1}_{J_1}\dots\delta^{I_l}_{J_{l+1}}\delta^{I_{l+1}}_{J_{l}}\dots\delta^{I_L}_{J_L}
\end{equation}
while $\mathcal{K}_{l,l+1}$ is the trace operator
\begin{equation}
\left(\mathcal{K}_{l,l+1}\right)^{I_1 I_2 \cdots I_L}_{J_1 J_2 \cdots J_L} %\delta^{I_1}_{J_1}\dots\delta^{I_l}_{J_l}\delta^{I_{l+1}}_{J_{l+1}}\dots\delta^{I_L}_{J_L}
=\delta^{I_1}_{J_1}\dots\delta^{I_l,I_{l+1}}\delta_{J_{l},J_{l+1}}\dots\delta^{I_L}_{J_L}
\end{equation}

Up to this point we have the Hamiltonian of an $SO(6)$ open spin chain which has been widely studied in the literature for different setups \cite{Berenstein:2005vf,DeWolfe:2004zt,Erler:2005nr}. The specificity of our spin chain will come from the boundary conditions, which are characterized by the remaining $H_{\mathrm{bdry}}$ piece. This piece comes from the contribution of the diagrams \ref{fig:Hc}-\ref{fig:Hf}.
Notice that those diagrams involve contractions of the first/last or last/first scalars of the chain with a scalar insertion in the loop. This is possible since in the set of operators we are considering we include the possibility of having $\Phi^4$ in the chain. Notice also that this contribution will vanish if the first (or last) scalar in the operator is not $\Phi^4$. While diagrams \ref{fig:Hc} and \ref{fig:Hf} do not contribute because they are not divergent, the net contribution to the two point function from \ref{fig:Hd} and \ref{fig:He} becomes
\begin{align}
2\times\left(\frac{\lambda}{8\pi^2}\right)^{L+1}\frac{\zeta^2}{\tau^{2L-2}}
& \left[\int _0^\tau d\tau_2\int _0^{\tau_2} d\tau_1
\frac{1}{\tau_1^{2-2\epsilon}(\tau-\tau_2)^{2-2\epsilon}}
\right]\nonumber\\
&\times\left(
\delta^{I_1}_4\delta_{J_1}^4\ \delta^{I_2}_{J_2}\dots\delta^{I_L}_{J_L}+
\delta^{I_L}_4\delta_{J_L}^4\ \delta^{I_1}_{J_1}\dots\delta^{I_{L-1}}_{J_{L-1}}
\right)
\end{align}
The integral is %(see appendix \ref{app:Integrals})
\begin{equation}
\int _0^\tau d\tau_2\int _0^{\tau_2} d\tau_1
\frac{1}{\tau_1^{2-2\epsilon}(\tau-\tau_2)^{2-2\epsilon}}
=-\frac{1}{\tau^2}\,\frac{1}{\epsilon}+\mathcal{O}(\epsilon^0)
\end{equation}
So we finally obtain
\begin{equation}
H_{\mathrm{bdry}}=\frac{\lambda\zeta^2}{8\pi^2}(\mathcal{Q}_4^{(1)}+\mathcal{Q}_4^{(L)})
\end{equation}
%
%such that we obtain for the renormalization constant
%\begin{equation}
%\mathcal{Z}_{\mathrm{bdry}}^{(1)}=\frac{\lambda}{16\pi^2}\frac{\zeta^2}{\epsilon}\left(\mathcal{Q}^{(1)}+\mathcal{Q}^{(L)}\right)
%\end{equation}
where the operators $\mathcal{Q}_4^{(1)}$ and $\mathcal{Q}_4^{(L)}$ act as the
identity when the corresponding site is occupied by a field $\Phi^4$
%project the flavour identity
\begin{equation}
\left(\mathcal{Q}_4^{(1)}\right)^{I_1 I_2 \cdots I_L}_{J_1 J_2 \cdots J_L}
%\delta^{I_1}_{J_1}\dots\delta^{I_L}_{J_L}
=\delta^{I_1}_4\delta_{J_1}^4\ \delta^{I_2}_{J_2}\dots\delta^{I_L}_{J_L},\qquad
\left(\mathcal{Q}_4^{(L)}\right)^{I_1 I_2 \cdots I_L}_{J_1 J_2 \cdots J_L}
 %\delta^{I_1}_{J_1}\dots\delta^{I_L}_{J_L}
 =\delta^{I_L}_4\delta_{J_L}^4\ \delta^{I_1}_{J_1}\dots\delta^{I_{L-1}}_{J_{L-1}}
\end{equation}

Therefore, the full one-loop dilatation operator for $SO(6)$ insertions of length $L$ is
\begin{equation}
\mathcal{D}^{(1)}(\zeta)=
\frac{\lambda}{8\pi^2}\left[
(\zeta^2-1)+\sum\limits_{l=1}^{L-1}\left(1-\mathcal{P}_{l,l+1}+\frac{1}{2}\mathcal{K}_{l,l+1}\right)+\zeta^2(\mathcal{Q}_4^{(1)}+\mathcal{Q}_4^{(L)})
\right]
\end{equation}

To verify this result with a simple computation we can apply the dilatation operator on length $L=1$ operators. { In this case there is no bulk term in $\mathcal{D}^{(1)}(\zeta)$ and the six possible single insertions are eigenstates. We consider $i\ne 4$ and obtain
\begin{align}
&O^{i}=\Phi^i \quad
& \to\quad &
\Delta=
1+\frac{(\zeta^2-1)}{8\pi^2}\lambda+\mathcal{O}(\lambda^2)
\nonumber\\
&O^{4}=\Phi^4 \quad
& \to\quad &
\Delta=
1+\frac{(3\zeta^2-1)}{8\pi^2}\lambda+\mathcal{O}(\lambda^2)
\end{align}
As expected, we recovered the result for the WML for $\zeta = \pm 1$, {\it i.e.} the $\Phi^i$ are protected
with dimension $\Delta=1$ while $\Phi^4$ is not with $\Delta=1+\tfrac{\lambda}{4\pi^2}+\mathcal{O}(\lambda^2)$ \cite{Alday:2007he}. On the other hand, as expected for the ordinary WL, for $\zeta=0$ the six scalar are on the same footing with dimension $\Delta=1-\tfrac{\lambda}{8\pi^2}+\mathcal{O}(\lambda^2)$ \cite{Alday:2007he}.
} It is interesting to notice that the dimension of the $\Phi^4$ field increases from its bare dimension for $|\zeta|>\tfrac{1}{\sqrt{3}}$ while it decreases for $|\zeta|<\tfrac{1}{\sqrt{3}}$.

Other examples are length $L=2$ operators, which can be constructed with relative ease to be eigenstates of $\mathcal{D}^{(1)}(\zeta)$. There are a total of $36$ independent operators and the brute force computation involves diagonalizing a $36$ by $36$ matrix. Consider $i,j,k,l\ne 4$, we have
\begin{align}
&\mbox{10 op:}\ \ O_A^{ij}=\Phi^i\Phi^j-\Phi^j\Phi^i \! & \to\  & \Delta=2+\frac{\left(\zeta^2+1\right)}{8\pi^2}\lambda+\mathcal{O}(\lambda^2)
\nonumber\\
&\mbox{14 op:}\ \ O_S^{ij}=\Phi^i\Phi^j+\Phi^j\Phi^i-\frac{2}{5}\delta^{ij}\delta_{kl}\Phi^k\Phi^l
\! & \to\  &  \Delta=2+\frac{\left(\zeta^2-1\right)}{8\pi^2}\lambda+\mathcal{O}(\lambda^2)
\nonumber\\
&\mbox{5 op:}\ \ O_A^{i}=\Phi^4\Phi^i-\Phi^i\Phi^4 \! & \to\  & \Delta=2+\frac{\left(2\zeta^2+1\right)}{8\pi^2}\lambda+\mathcal{O}(\lambda^2)
\nonumber\\
&\mbox{5 op:}\ \ O_S^{i}=\Phi^4\Phi^i+\Phi^i\Phi^4  \! & \to\  &  \Delta=2+\frac{\left(2\zeta^2-1\right)}{8\pi^2}\lambda+\mathcal{O}(\lambda^2)\nonumber\\
&\mbox{2 op:}\ \ O_{\pm}=\delta_{ij}\Phi^i\Phi^j+2(\zeta^2-1\pm A(\zeta))\Phi^4\Phi^4
\! & \to\  &  \Delta=2+\frac{\lambda}{8\pi^2}\left(\tfrac{1}{2}+2\zeta^2\pm A(\zeta)\right) +\mathcal{O}(\lambda^2)\nonumber
\end{align}
where $A(\zeta)=\sqrt{(\zeta^2-1)^2+\tfrac{5}{4}}$. It is clear that the problem becomes increasingly difficult when the length of the operator increases.

\section{Bethe ansatz and boundary Yang-Baxter equation}

To study the integrability of this spectral problem we ask if it can be solved with a Bethe ansatz. In order to do so, we first define a vacuum $|0\rangle$ associated to the operator $Z^L$ with $Z=(\Phi^5+i\Phi^6)/\sqrt{2}$. Since $Z^L$ is symmetric, traceless and contains no $\Phi^4$ field it has $H^{(1)}(\zeta)|0\rangle=E_0 |0\rangle$. From now on we shall use lower case Latin letters from the beginning of the alphabet $a,b,c,\dots$ for flavour indexes that range
from $1$ to $4$. While one could work using complex fields, besides $Z$ we choose to keep real fields since our boundary conditions will be non-chiral.

On top of the vacuum we shall put impurities with flavours from $1$ to $4$. We associate the operators to spin chain states as follows
\begin{equation}
Z\dots Z\underbrace{\Phi_a}_l Z\dots Z\to |l\rangle_a,\qquad
Z\dots Z\underbrace{\Phi_a}_{l_1} Z\dots Z\underbrace{\Phi_b}_{l_2}Z\dots Z\to |l_1,l_2\rangle_{a b}
% Z\dots Z\underbrace{\bar Z}_l Z\dots Z\to |l\rangle_{\bar Z}
\end{equation}
and so on. We shall eventually use the state $|l\rangle_{\bar Z}$ which is given by a $\bar Z$ excitation in the $l$-th place since once we have at least two neighbor excitations the trace $\mathcal{K}$ will create them as in
\begin{equation}
\mathcal{K}_{l,l+1}|l,l\!+\!1\rangle_{ab}=
\delta_{ab}\left(\delta^{cd}|l,l\!+\!1\rangle_{cd}+|l\rangle_{\bar Z}+|l\!+\!1\rangle_{\bar Z}\right)
\end{equation}
If the chain had no boundaries, one magnon excitations of the form
\begin{equation}
|\psi(p)\rangle_a=\sum\limits_{l}e^{i p l}|l\rangle_a
\end{equation}
would be eigenstates of $H^{(1)}(\zeta)$ with eigenvalue $E_0 + \varepsilon(p)$, where $\varepsilon(p)=\tfrac{\lambda}{2\pi^2}\sin^2\tfrac{p}{2}$.

In the presence of boundaries, eigenstates are obtained as superpositions of left moving and right moving magnons using the reflection matrix $R_a^b(p)$.
\begin{equation}
|\Psi(p)\rangle_a=|\psi(p)\rangle_a+R_a^b(p)|\psi(-p)\rangle_b
%,\quad\mbox{with}\quad
%|\psi(p)\rangle_a=\sum\limits_{l=1}^{L}e^{i p l}|l\rangle_a
\end{equation}
This state is an eigenstate of $H^{(1)}(\zeta)$ for a specific reflection matrix. From the left boundary we find
\begin{equation}
R_a^b=R_T\left(\delta_a^b-\delta_a^4 \delta_4^b\right)+R_{\parallel}\delta_a^4 \delta_4^b,
\end{equation}
this is, a different reflection matrix for `transverse' (with flavours $1, 2$ and $3$) and `parallel' (with flavour $4$) excitations. These are given by
\begin{equation}
R_T(p)=e^{i p}\qquad\mbox{and}\qquad R_{\parallel}(p)=-\frac{1+e^{ip}(\zeta^2-1)}{1+e^{-ip}(\zeta^2-1)}
\end{equation}
Notice that $R_{\parallel}=-1$ for the WML while $R_{\parallel}=R_{T}$ for the ordinary WL. The right boundary on the other hand provides us with the quantization conditions for the momenta. For transverse excitations we find $\left(R_T\right)^2=e^{2i p (L+1)}$ while for parallel ones $\left(R_\parallel\right)^2=e^{2i p (L+1)}$.

In order to obtain the scattering matrix for two magnon excitations we forget about the boundaries and focus on the bulk Hamiltonian. This one loop scattering matrix was obtained in \cite{Berenstein:2005vf} but there they were using complex fields and we now reformulate it in terms of real fields. Our coordinate Bethe ansatz for two excitations is
\begin{equation}
\label{CoordBethe2}
|\Psi(p_1, p_2)\rangle_{ab} = |\psi(p_1, p_2)\rangle_{ab} + S_{ab}^{cd}(p_1,p_2) |\psi(p_2, p_1)\rangle_{cd} + \sigma_{ab}(p_1,p_2) |\gamma(p_1+p_2)\rangle
\end{equation}
with
\begin{equation}
|\psi(p_1, p_2)\rangle_{ab} = \sum\limits_{l_2>l_1}e^{i(p_1 l_1+p_2 l_2)} |l_1,l_2\rangle_{ab}\quad\mbox{and}\quad
 |\gamma(p_1+p_2)\rangle = \sum\limits_l e^{i(p_1+p_2)l} |l\rangle_{\Bar{Z}}
\end{equation}
In other words, a term for the incoming magnons, a term for the scattered magnons and a term which looks as a decay to $\bar Z$ excitations with momenta $p_1+p_2$. This extra term is needed due to the presence in the Hamiltonian of trace $\mathcal{K}$ operators.

Applying $H_{\mathrm{bulk}}$ to (\ref{CoordBethe2}) we find it is an eigenstate with eigenvalue $\varepsilon(p_1)+\varepsilon(p_2)$ as long as the following equations are satisfied
\begin{align}
\label{EquationDefiningSMatrix}
\Omega_{ab}^{cd}(p_1,p_2)+S_{ab}^{ef}(p_1,p_2)\Omega_{ef}^{cd}(p_2,p_1)+\sigma_{ab}(p_1,p_2)\delta^{cd}g(p_1,p_2)=0&\\
 e^{-ip_1}g(p_1,p_2)\delta_{ab} +  e^{-ip_2}g(p_1,p_2)S_{ab}^{cd}(p_1,p_2)\delta_{cd} + \sigma_{ab}(p_1,p_2)f(p_1,p_2)=0&
\end{align}
where
\begin{align}
& \Omega_{ab}^{cd}(p_1,p_2)=(e^{ip_1+ip_2}-e^{ip_2}+1)\delta_a^c\delta_b^d
-e^{ip_2}\ \delta_a^d\delta_b^c+\frac{e^{ip_2}}{2}\delta_{ab}\delta^{cd}\nonumber\\
& g(p_1,p_1)=\frac{1+e^{ip_1+ip_2}}{2},\qquad
f(p_1,p_2)=g(p_1,p_2)(2e^{-ip_1}+2e^{-ip_1}-e^{-ip_1-ip_2}-1)
\end{align}
The solution for the scattering matrix of (\ref{EquationDefiningSMatrix}) is given by
%\footnote{We use symmetric and antisymmetric combinations of deltas defined by $\delta^{cd}_{\{ab\}}=\tfrac{1}{2}\left(\delta_a^c\delta_b^d+\delta_a^d\delta_b^c\right)$ and $\delta^{cd}_{[ab]}=\tfrac{1}{2}\left(\delta_a^c\delta_b^d-\delta_a^d\delta_b^c\right)$}
\begin{equation}\label{SMatrixNonSymmetric}
S_{ab}^{cd}(p_1,p_2) = \frac{S(p_1,p_2)-1}{2}\delta_a^c\delta_b^d+\frac{S(p_1,p_2)+1}{2}\delta_a^d\delta_b^c+\frac{S(p_2,p_1)-S(p_1,p_2)}{4}\delta_{ab}\delta^{cd}
\end{equation}
where
\begin{equation}
\label{Ssu2}
S(p_1,p_2)=-\frac{e^{ip_1+ip_2}-2e^{ip_2}+1}{e^{ip_1+ip_2}-2e^{ip_1}+1}
\end{equation}
is the $SU(2)$ scattering phase. It is interesting to rewrite (\ref{SMatrixNonSymmetric}) in terms of $SO(4)$ projectors
\begin{equation}\label{SMatrixSymmetric}
\mathbb{S}(p_1,p_2)=S_{\mathfrak{su}(2)}\Pi_S\ +S_{\mathfrak{su}(1|1)}\Pi_A\ +S_{\mathfrak{sl}(2)}\Pi_T
\end{equation}
where $\Pi_S$ is the symmetric traceless projector, $\Pi_A$ the antisymmetric projector and $\Pi_T$ the trace projector. They are given in components by
\begin{equation}
(\Pi_S)_{ab}^{cd}=\tfrac{1}{2}\delta_a^c\delta_b^d+\tfrac{1}{2}\delta_a^d\delta_b^c-\tfrac{1}{4}\delta_{ab}\delta^{cd}, \quad
(\Pi_A)_{ab}^{cd}=\tfrac{1}{2}\delta_a^c\delta_b^d-\tfrac{1}{2}\delta_a^d\delta_b^c,\quad
(\Pi_T)_{ab}^{cd}=\tfrac{1}{4}\delta_{ab}\delta^{cd}
\end{equation}
We defined $S_{\mathfrak{su}(2)}=S(p_1,p_2)$, $S_{\mathfrak{su}(1|1)}=-1$ and $S_{\mathfrak{sl}(2)}=S(p_2,p_1)$ \cite{Staudacher:2004tk}. The $S$-matrix as written in (\ref{SMatrixSymmetric}) admits a very nice interpretation. It says that a pair of magnons scatter with a simple $SU(2)$ phase if they are in a symmetric traceless representation of $SO(4)$, with an $SU(1|1)$ phase if they are in the antisymmetric representation of $SO(4)$ and with an $SL(2)$ phase if they are in singlet trace of $SO(4)$.

The solution for the coupling to the $\bar Z$ excitations
%is more difficult to interpret\footnote{\color{red}\bf NDR: en realidad ahora me di cuenta de una manera precisa de interpretarlo, pero no se si vale la pena escribirlo}. It
is given by
\begin{equation}
\sigma_{ab}(p_1,p_2)=-\frac{1}{2}(1+S(p_2,p_1))\delta_{ab}
\end{equation}
The necessary and sufficient condition for the bulk hamiltonian to be integrable is that scattering of three or more magnons is factorized as products of two particle scattering matrices. A consistency condition for this factorization is the  Yang-Baxter equation. In terms of $S$ and $\sigma$ this implies the constraints which appear for a three-excitations Bethe ansatz
\begin{align}
S_{ab}^{df}(p_1,p_2)S_{fc}^{ei}(p_1,p_3)S_{de}^{gh}(p_2,p_3)= &
S_{bc}^{fd}(p_2,p_3)S_{af}^{ge}(p_1,p_3)S_{ed}^{hi}(p_1,p_2)\nonumber\\
\sigma_{ab}(p_1,p_2)= & S_{ab}^{cd}(p_1,p_2)\sigma_{cd}(p_2,p_1)
\end{align}
and it can be verified explicitly that they are satisfied.

It will be useful to consider the scattering matrix in terms of four different choices for the incoming and outgoing magnons. Let $a\ne b$, we could have equal or different flavour incoming or outgoing magnons (no sum implied for $a$ and $b$)
\begin{align}
& S_{ab}^{ab} (p_1,p_2) \equiv S^- (p_1,p_2)=\frac{1}{2}(S(p_1,p_2) - 1) \quad S_{ab}^{ba} (p_1,p_2) \equiv S^+(p_1,p_2)=\frac{1}{2}(S(p_1,p_2) + 1) \nonumber\\
& S_{aa}^{bb} (p_1,p_2) \equiv S_D (p_1,p_2)= \frac{1}{4}(S(p_2,p_1) - S(p_1,p_2)) \nonumber\\
& S_{aa}^{aa} (p_1,p_2) \equiv S_I (p_1,p_2)= \frac{1}{4}(S(p_2,p_1) + 3S(p_1,p_2))
\end{align}

The analogue of the bulk Yang-Baxter equation for the boundary exists. For the problem with boundaries, a consistency condition for integrability requires that the scattering matrix and the reflection matrix obey the equation
\begin{equation}\label{BoundaryYB}
R_a^c(p_1)S_{cb}^{de}(-p_1,p_2) R_d^f (p_2) S_{fe}^{gh}(-p_2, -p_1) = S_{ab}^{cd}(p_1,p_2)R_c^e(p_2)S_{ed}^{fh}(-p_2,p_1) R_f^g(p_1)
\end{equation}

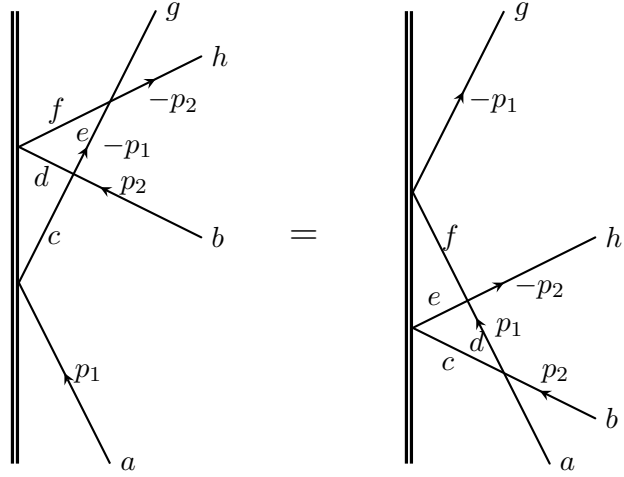
\begin{figure}[!ht]
    \centering
\begin{subfigure}[b]{0.22\textwidth}
\begin{tikzpicture}[scale=0.6,>=stealth]
\draw[very thick] (-0.03,-5)--(-0.03,5);
\draw[very thick] (-0.14,-5)--(-.14,5);
\draw[thick,->](2,-5)--(1,-3);
\draw[thick,cap=round](1.05,-3.1)-- (0,-1);
\draw[thick,->] (0,-1)--(1.5,2);
\draw[thick,cap=round] (1.45,1.9)--(3,5);
\draw[thick,cap=round](2.9,3.45)--(4,4);
\draw[thick,->](0,2)--(3,3.5);
\draw[thick,-<](0,2)--(2,1);
\draw[thick,cap=round](1.9,1.05)--(4,0);
\draw (2,-5) node[right] {$a$};
\draw (4,0) node[right] {$b$};
\draw (0.4,0) node[right] {$c$};
\draw (0.1,1.35) node[right] {$d$};
\draw (1.4,1.9) node[above] {$e$};
\draw (0.4,2.8) node[right] {$f$};
\draw (3,5) node[right] {$g$};
\draw (4,4) node[right] {$h$};
\draw (1,-3) node[right] {$p_1$};
\draw (2,1.15) node[right] {$p_2$};
\draw (2.6,3.) node[right] {$-p_2$};
\draw (1.6,2.) node[right] {$-p_1$};
\draw (5.7,0) node[right] {\Large$=$};
\end{tikzpicture}
\end{subfigure}
\hspace{1.6cm}
\begin{subfigure}[b]{0.22\textwidth}
\begin{tikzpicture}[scale=0.6,>=stealth]
\draw[very thick] (-0.03,-5)--(-0.03,5);
\draw[very thick] (-0.14,-5)--(-.14,5);
\draw[thick,-<](2,5)--(1,3);
\draw[thick,cap=round](1.05,3.1)-- (0,1);
\draw[thick,-<] (0,1)--(1.5,-2);
\draw[thick,cap=round] (1.45,-1.9)--(3,-5);
\draw[thick,cap=round](2.9,-3.45)--(4,-4);
\draw[thick,-<](0,-2)--(3,-3.5);
\draw[thick,->](0,-2)--(2,-1);
\draw[thick,cap=round](1.9,-1.05)--(4,0);
\draw (3,-5) node[right] {$a$};
\draw (4,-4) node[right] {$b$};
\draw (0.4,-2.8) node[right] {$c$};
\draw (1.4,-1.85) node[below] {$d$};
\draw (0.4,0) node[right] {$f$};
\draw (0.1,-1.35) node[right] {$e$};
\draw (2,5) node[right] {$g$};
\draw (4,0) node[right] {$h$};
\draw (1,3) node[right] {$-p_1$};
\draw (2,-1.1) node[right] {$-p_2$};
\draw (2.6,-3.) node[right] {$p_2$};
\draw (1.6,-2.) node[right] {$p_1$};
\end{tikzpicture}
\end{subfigure}
 \caption{Graphical representation of the boundary Yang-Baxter equation. $a,b,c,d={ 1},\dots,4$ are flavour indices while $p_1$ and $p_2$ are the magnon momenta.}
  \label{fig:BoundaryYB}
\end{figure}

In figure \ref{fig:BoundaryYB} we have a graphical representation of this equation. Incoming magnons with flavours $a$ and $b$ can scatter and bounce in the boundary in two different ways which have to be equivalent. One can check that (\ref{BoundaryYB}) is trivial for most of its components. In fact, if the incoming particles have different flavours, such that they involve the $S^{+}$ and $S^{-}$ components of the $S$-matrix, (\ref{BoundaryYB}) is trivially true. The only non-trivial components of equation (\ref{BoundaryYB}) are those in which the incoming particles are parallel excitations ($a=b=4$) and the outcoming ones are transverse excitations ($g=h=1,2,3$) and vice versa. In this case the equation reads
\begin{align}\label{BoundaryYBNonTrivial}
&R_{\parallel}(p_1)\left[S_I(-p_1,p_2)S_D(-p_2,-p_1)R_\parallel(p_2)+R_T(p_2)S_D(-p_1,p_2)
S_I(-p_1,-p_2)\right]\nonumber\\
&=R_T(p_1)\left[S_I(p_1,p_2)R_\parallel(p_2) S_D (-p_2,p_1) + S_D(p_1,p_2)R_T(p_2)
S_I(p_1,-p_2)\right]
\end{align}
Taking the difference between the LHS and the RHS of (\ref{BoundaryYBNonTrivial}), substituting for $R_\parallel, R_T, S_D$ and $S_I$ and noting $z_i=e^{ip_i}$ we get
\begin{equation}
0=(S(z_1,z_2)\!+\!S(z_2,z_1))
\frac{z_1 z_2(1-z_1^2)(1-z_2^2)(1-z_1 z_2)}{(z_1+z_2-2)(2z_1 z_2-z_1-z_2)}\
\frac{\zeta^2(\zeta^2-1)}{(z_1+\zeta^2-1)(z_2+\zeta^2-1)}
\end{equation}
and therefore the complete boundary Yang-Baxter equation (\ref{BoundaryYB}) is fulfilled only if
\begin{equation}
\zeta=0\qquad \mbox{or}\qquad\zeta=\pm 1
\end{equation}
which correspond to the ordinary WL or the WML respectively. Therefore we can claim that there is no integrability for arbitrary values of $\zeta$ and that there is 1-loop integrability in the $SO(6)$ sector for the ordinary WL and the WML only. For those cases the transverse and parallel reflection matrices become
\begin{align}
& R_T(p)=e^{ip},\qquad R_{\parallel}(p)=e^{ip}\qquad\qquad &\mbox{WL}\nonumber\\
& R_T(p)=e^{ip},\qquad R_{\parallel}(p)=-1\qquad\qquad &\mbox{WML}
\end{align}

\section{One loop dilatation operator in the $SU(2|3)$ sector}
In order to further test the integrability of the boundary condition set by the ordinary WL with $\zeta=0$,
we now consider the insertion of operators made of 3 complex scalar fields $\phi^a$ and 2 complex fermion fields $\chi^\alpha$. The three complex scalars are in a $\bf 3$ of $SU(3)\subset SU(4)_R$ while the fermion $\chi$ is a singlet of that symmetry. On the other hand the fermion is a $\bf 2$ of an $SU(2)$ of the Lorentz group while the scalars are obviously invariant. It should be noticed that this sector makes sense as a closed sector in single trace operators while when inserted on the Wilson loop it will be closed only for $\zeta=0$ since the scalar in the general Wilson loop with $\zeta\ne 0$ is real and mixes $\phi^a$ with $\bar\phi^a$.

More precisely we consider insertions  of the form $O^{A_1...A_L}=W^{A_1}...W^{A_L}$ where indices $A$ run from 1 to 5 and
\begin{equation}
W^1 = \phi^1,\quad W^2 = \phi^2,\quad W^3 = \phi^3,\quad
W^4 = \chi^1,\quad W^5 = \chi^2,\quad
\end{equation}

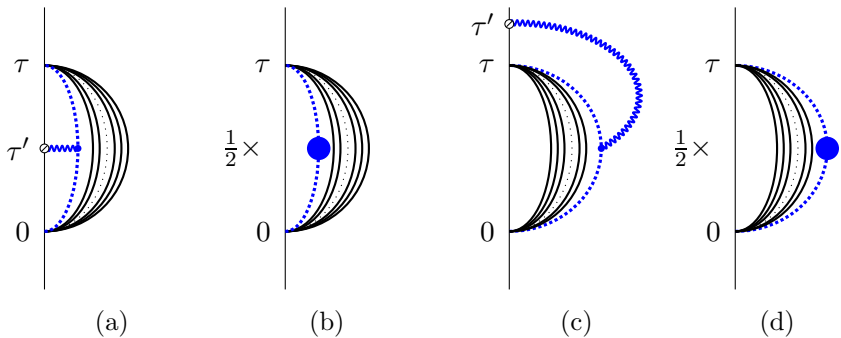
\begin{figure}[!ht]
    \centering
\begin{subfigure}[b]{0.2\textwidth}
\begin{tikzpicture}[scale=1.1]
\draw  (0,-1.7) -- (0,1.7);
\draw[thick] (0,-1) arc (-90:90: 1 and 1) ;
\draw[thick] (0,-1) arc (-90:90: 0.92 and 1) ;
\draw[thick] (0,-1) arc (-90:90: 0.84 and 1) ;
\draw[dotted] (0,-1) arc (-90:90: 0.75 and 1) ;
\draw[thick] (0,-1) arc (-90:90: 0.66 and 1) ;
\draw[thick] (0,-1) arc (-90:90: 0.58 and 1) ;
\draw[very thick, densely dotted, blue] (0,-1) arc (-90:90: 0.4 and 1) ;
\draw[blue, thick,decoration={snake,amplitude=1pt, segment length=2pt}, decorate]  (0,0) -- (0.4,0);
\fill[blue] (0.4,0) circle (1.2pt);
\fill[white] (0,0) circle (1.5pt);
\draw (0-0.033,-0.033)--(+0.033,+0.033);
\draw (0,0) circle (1.5pt);
\draw (-0.05,-1) node[left] {$0$};
\draw (-0.05,1) node[left] {$\tau$};
\draw (-0.05,0) node[left] {$\tau'$};
\end{tikzpicture}
        \caption{}
        \label{fig2:Ha}
    \end{subfigure}\!\!\!\!\!
    \begin{subfigure}[b]{0.2\textwidth}
\begin{tikzpicture}[scale=1.1]
\draw  (0,-1.7) -- (0,1.7);
\draw[thick] (0,-1) arc (-90:90: 1 and 1) ;
\draw[thick] (0,-1) arc (-90:90: 0.92 and 1) ;
\draw[thick] (0,-1) arc (-90:90: 0.84 and 1) ;
\draw[dotted] (0,-1) arc (-90:90: 0.75 and 1) ;
\draw[thick] (0,-1) arc (-90:90: 0.66 and 1) ;
\draw[thick] (0,-1) arc (-90:90: 0.58 and 1) ;
\draw[very thick, densely dotted, blue] (0,-1) arc (-90:90: 0.4 and 1) ;
\fill[blue] (0.4,0) circle (4pt);
\draw (-0.05,-1) node[left] {$0$};
\draw (-0.05,1) node[left] {$\tau$};
\draw (-0.15,0) node[left] {$\tfrac{1}{2}\times$};
\end{tikzpicture}
        \caption{}
        \label{fig2:Hb}
    \end{subfigure}\ \
        \begin{subfigure}[b]{0.2\textwidth}
\begin{tikzpicture}[scale=1.1]
\draw  (0,-1.7) -- (0,1.7);
\draw[blue, very thick, densely dotted] (0,-1) arc (-90:90: 1.1 and 1) ;
\draw[thick] (0,-1) arc (-90:90: 0.92 and 1) ;
\draw[thick] (0,-1) arc (-90:90: 0.84 and 1) ;
\draw[dotted] (0,-1) arc (-90:90: 0.75 and 1) ;
\draw[thick] (0,-1) arc (-90:90: 0.66 and 1) ;
\draw[thick] (0,-1) arc (-90:90: 0.58 and 1) ;
\draw[thick] (0,-1) arc (-90:90: 0.5 and 1) ;
\draw (-0.05,-1) node[left] {$0$};
\draw (-0.05,1) node[left] {$\tau$};
\draw (-0.05,1.5) node[left] {$\tau'$};
\fill[blue] (1.1,0) circle (1.2pt);
\draw[blue, thick,decoration={snake,amplitude=1pt, segment length=2pt}, decorate]  (1.1,0) arc (-45:93: 1.5 and 0.88);
\fill[white] (0,1.5) circle (1.5pt);
\draw (0-0.033,1.5-0.033)--(0+0.033,1.5+0.033);
\draw (0,1.5) circle (1.5pt);
\end{tikzpicture}
        \caption{}
        \label{fig2:Hc}
    \end{subfigure}\!\!\!\!\!\!\!\!
    \begin{subfigure}[b]{0.2\textwidth}
\begin{tikzpicture}[scale=1.1]
\draw  (0,-1.7) -- (0,1.7);
\draw[blue, very thick, densely dotted] (0,-1)
arc (-90:90: 1.1 and 1) ;
\draw[thick] (0,-1) arc (-90:90: 0.92 and 1) ;
\draw[thick] (0,-1) arc (-90:90: 0.84 and 1) ;
\draw[dotted] (0,-1) arc (-90:90: 0.75 and 1) ;
\draw[thick] (0,-1) arc (-90:90: 0.66 and 1) ;
\draw[thick] (0,-1) arc (-90:90: 0.58 and 1) ;
\draw[thick] (0,-1) arc (-90:90: 0.5 and 1) ;
\draw (-0.05,-1) node[left] {$0$};
\draw (-0.05,1) node[left] {$\tau$};
\fill[blue] (1.1,0) circle (4pt);
\draw (-0.15,0) node[left] {$\tfrac{1}{2}\times$};
\end{tikzpicture}
        \caption{}
        \label{fig2:Hd}
    \end{subfigure}
    \caption{Feynman diagrams contributing to $H_{\mathrm{bdry}}$. We represent fermion and gluon propagators by dotted and wiggled lines. Blue bullets represent 1-loop self-energy corrections to the propagators.}
    \label{fig:ferBT}
\end{figure}

As in the $SO(6)$ sector, there will be three types of contributions to the 1-loop dilatation operator
\begin{equation}
\mathcal{D}^{(1)}=E_0 +H_{\mathrm{bulk}} +H_{\mathrm{bdry}},
\end{equation}
where the contribution $E_0$ is the same as before, $H_{\mathrm{bulk}}$ is the 1-loop $SU(2|3)$ spin chain Hamiltonian \cite{Beisert:2003ys} and $H_{\mathrm{bdry}}$ comes from the interaction between leftmost and rightmost fields of the insertions with the Wilson loop. As seen in the previous section, this interaction is vanishing for $\zeta = 0$ and scalar fields occupying the sites 1 and $L$. So, it remains to determine the boundary terms for the case of fermion fields occupying the sites 1 and $L$. Such contribution can be inferred from the diagrams in figure \ref{fig:ferBT} and will lead to diagonal boundary terms.

Thus, the resulting 1-loop dilatation operator would be of the form
\begin{equation}
\mathcal{D}^{(1)} = 2g^2\left[
-1+\sum\limits_{l=1}^{L-1}\left(1-\Pi_{l,l+1}\right)
+\alpha(\mathcal{Q}_F^{(1)}+\mathcal{Q}_F^{(L)})
\right]+2g^3 H^{(3)},
\label{1looDsu23}
\end{equation}
where $g^2=\tfrac{\lambda}{16\pi^2}$,
\begin{equation}\label{H3}
H^{(3)}=-  e^{i\beta}\epsilon_{\alpha\beta}\epsilon^{ijk}
\Big\{\resizebox{0.6cm}{!}{\begin{tabular}{c}
$\!\!\alpha\beta\!\!$\\
$\!\!i j k\!\!$
\end{tabular}}\Big\}
- e^{-i\beta}\epsilon^{\alpha\beta}\epsilon_{ijk}
\Big\{\resizebox{0.6cm}{!}{\begin{tabular}{c}
$\!\!i j k\!\!$\\
$\!\!\alpha\beta\!\!$
\end{tabular}}\Big\}
\end{equation}
and $\Pi_{l,l+1}$ is the graded permutation operator, {\it i.e.} it permutes the fields at sites $l$ and $l\!+\!1$ with an additional sign of both fields are fermionic. The symbol $\Big\{
^{A_1 A_2\cdots A_n}
_{B_1 B_2\cdots B_m}
\Big\}$ represents the action of replacing a sequence of consecutive fields $W^{A_1}W^{A_2}\cdots W^{A_n}$ anywhere along the insertion by the fields $W^{B_1}W^{B_2}\cdots W^{B_m}$. Thus, (\ref{H3})
is a term that changes the length of the spin chain, since it converts three bosons into two fermions and vice versa.

The boundary operators $\mathcal{Q}_F^{(1)}$ and $\mathcal{Q}_F^{(L)}$  are now
\begin{equation}
\left( \mathcal{Q}^{(1)}_F \right)^{A_1 A_2 \cdots A_L}_{B_1 B_2 \cdots B_L}
=\delta^{A_1}_\alpha\delta_{J_1}^\alpha\ \delta^{A_2}_{B_2}\dots\delta^{A_L}_{B_L},\qquad
\left( \mathcal{Q}^{(L)}_F\right)^{A_1 A_2 \cdots A_L}_{B_1 B_2 \cdots B_L}
 =\delta^{A_L}_\alpha\delta_{B_L}^\alpha\ \delta^{A_1}_{B_1}\dots\delta^{A_{L-1}}_{B_{L-1}}
\end{equation}

A few comments are in order about the term $H^{(3)}$.  Although its action is immaterial for the spectrum at 1-loop order, it does modify the Bethe ansatz wave-functions in an order that is intermediate between 1-loop and 2-loop. As a consequence of that, the Yang-Baxter equations will have an expansion
including as well an order intermediate between 1-loop and 2-loop.

Concerning the boundary terms, we should emphasize that $\alpha$ is not a free parameter but uniquely fixed by the diagrams in figure \ref{fig:ferBT}. Diagrams \ref{fig2:Ha} and \ref{fig2:Hc} give
\begin{equation}
\mbox{\ref{fig2:Ha}} + \mbox{\ref{fig2:Hc}}= T_{\alpha}^{\dot\alpha}(\tau)\frac{g^2}{\epsilon}(\mathcal{Q}^{(1)}_F+\mathcal{Q}^{(L)}_F)
\end{equation}
where $T_{\alpha}^{\dot\alpha}(\tau)=\tfrac{iN}{4\pi^2}\frac{(\sigma^0)_{\alpha}^{\dot\alpha}}{\tau^3}$ is the tree-level result (for more details see Appendix).
The other contribution is half the 1-loop correction to a fermionic propagator each time a fermion is at the edge of the operator next to the Wilson loop as shown in \ref{fig2:Hb} and \ref{fig2:Hd}. The other half is used to construct $H_{\mathrm{bulk}}$. They give
\begin{equation}
\mbox{\ref{fig2:Hb}} + \mbox{\ref{fig2:Hd}}=-\frac{1}{2} T_{\alpha}^{\dot\alpha}(\tau)\frac{g^2}{\epsilon}({ 1+3})(\mathcal{Q}^{(1)}_F+\mathcal{Q}^{(L)}_F)
\end{equation}
where the `$1$' in the sum comes from the gluon correction of the propagator while the `$3$' from the scalar triplet correction. Collecting both contributions we get
\begin{equation}\label{HbdrySU23}
H_{\mathrm{bdry}} = 2g^2 \frac12(\mathcal{Q}_F^{(1)}+\mathcal{Q}_F^{(L)})
\end{equation}
In other words, the 1-loop dilatation operator is the one given in (\ref{1looDsu23}) with $\alpha = \frac12$.

As before, to diagonalize the dilatation operator we start by identifying the vacuum, which is again a chain of complex scalars of the same type: ${(\phi^3)}^L$. In this case, the magnon excitations that would propagate on top of it would transform in the fundamental representation of the residual symmetry $SU(2|2)$.

In the presence of a boundary we determine the reflection matrix
demanding that
\begin{equation}
|\Psi(p)\rangle_A = \sum\limits_{l=1}^{L}\left( e^{i p l}  |l\rangle_A +  e^{-i p l} R_A^B(p) |l\rangle_B \right)
\label{1magonwf}
\end{equation}
is an eigenstate of  $\mathcal{D}^{(1)}$. We find that $R_A^B(p)$ is diagonal with
\begin{equation}
R_a^b = R_\phi(p) \delta_a^{b} = e^{ip} \delta_a^{b},\quad \qquad
R_\alpha^\beta = R_\chi(p) \delta_\alpha^{\beta} = -\frac{1+e^{ip}(\alpha-1)}{1+e^{-ip}(\alpha-1)} \delta_\alpha^{\beta}
\label{RMatrixsu23}
\end{equation}

Although we have computed that $\alpha=\frac12$, we keep as if it were unspecified in the boundary scattering phase for fermions.
As we will soon see, the actual value of $\alpha$ turns out to be crucial for integrability to persist.

Let us now turn to the determination of the bulk scattering matrix. It is a well-known fact that beyond the 1-loop order the Bethe ansatz for ${\cal N} = 4$ SYM works only asymptotically. Strictly speaking the intermediate order term $H^{(3)}$ is beyond 1-loop and therefore the two-magnon wave-functions
\begin{equation}
|\Psi(p_1,p_2)\rangle_{AB}=|\psi(p_1,p_2)\rangle_{AB}+S_{AB}^{CD}(p_1,p_2)|\tilde\psi(p_2,p_1)\rangle_{CD}.
\label{an1}
\end{equation}
should incorporate some contact terms. Up to the order in (\ref{1looDsu23}), it is enough to correct the standard wave functions with contact terms in the following way
\begin{align}
|\psi(p_1,p_2)\rangle_{AB} &=
\sum\limits_{l_1<l_2}\left(
\delta_A^C\delta_B^D+g^{l_2-l_1}\,{F^{(1)}}_{AB}^{CD}
\right)e^{i l_1 p_1+i l_2 p_2}|l_1,l_2\rangle_{CD}
\label{an2}
\\
|\tilde\psi(p_2,p_1)\rangle_{AB} &=
\sum\limits_{l_1<l_2}\left(
\delta_A^C\delta_B^D+g^{l_2-l_1}\,{G^{(1)}}_{AB}^{CD}
\right)e^{i l_1 p_2+i l_2 p_1}|l_1,l_2\rangle_{CD}
\label{an3}
\end{align}
In accordance with (\ref{1looDsu23}) and the ansatz (\ref{an1})-(\ref{an3}), the $S$-matrix also has to be expanded
\begin{equation}
S_{AB}^{CD}={S^{(0)}}_{AB}^{CD}+g\,{S^{(1)}}_{AB}^{CD}+\mathcal{O}(g^2),
\end{equation}
and taking into account the contributions from the perturbative corrections of the $S$-matrix and the contact term corrections, the wave functions can be re-arranged in the form of the following pertubative expansion
\begin{equation}
|\Psi(p_1,p_2)\rangle_{AB}=|\Psi(p_1,p_2)\rangle_{AB}^{(0)}+g\,
|\Psi(p_1,p_2)\rangle_{AB}^{(1)}+\mathcal{O}(g^2)
\end{equation}

Since the eigenvalues of the eigenstates are expanded only in even powers of $g$
\begin{equation}
E=2 g^2 E^{(2)}+\mathcal{O}(g^4)
\end{equation}
where $E^{(2)}=4\sin^2\tfrac{p_1}{2}+4\sin^2\tfrac{p_2}{2}-1$, we obtain the following conditions from the first two orders of the eigenvalue equation
\begin{align}
H^{(2)} |\Psi(p_1,p_2)\rangle_{AB}^{(0)}
& = E^{(2)}|\Psi(p_1,p_2)\rangle_{AB}^{(0)}
\label{eieq1}
\\
H^{(3)} |\Psi(p_1,p_2)\rangle_{AB}^{(0)} + H^{(2)} |\Psi(p_1,p_2)\rangle_{AB}^{(1)}
& = E^{(2)}|\Psi(p_1,p_2)\rangle_{AB}^{(1)}
\label{eieq2}
\end{align}

The condition (\ref{eieq1}) implies that the leading order components of the $S$-matrix are
\begin{align}
{S^{(0)}}_{ab}^{cd}(p_1,p_2) &=\frac{S(p_1,p_2)-1}{2}\delta_a^c\delta_b^d+\frac{S(p_1,p_2)+1}{2}\delta_a^d\delta_b^c
\nonumber\\
{S^{(0)}}_{\alpha\beta}^{\gamma\delta}(p_1,p_2) &=\frac{S(p_1,p_2)-1}{2}\delta_\alpha^\gamma\delta_\beta^\delta-\frac{S(p_1,p_2)+1}{2}\delta_\alpha^\delta
\delta_\beta^\gamma
\nonumber\\
{S^{(0)}}_{a\alpha}^{b\beta}(p_1,p_2) &=S_{\alpha a}^{\beta b}(p_1,p_2)
=\frac{S(p_1,p_2)-1}{2}\delta_a^b\delta_\alpha^\beta
\nonumber\\
{S^{(0)}}_{a\alpha}^{\beta b}(p_1,p_2) &=S_{\alpha a}^{b\beta}(p_1,p_2)
=\frac{S(p_1,p_2)+1}{2}\delta_a^b\delta_\alpha^\beta
\label{SMatrixsu23}
\end{align}
where $S(p_1,p_2)$ is $SU(2)$ scattering phase given by (\ref{Ssu2}) and all other components vanishing.

The condition (\ref{eieq2}) on the other hand is non-trivial only for a two-scalar or a two-fermion state and can be satisfied establishing a linear relation between the contact factors ${F^{(1)}}_{AB}^{CD}$ and ${G^{(1)}}_{AB}^{CD}$
\begin{equation}
z_2\, {F^{(1)}}_{\alpha\beta}^{ij}+z_1\, {S^{(0)}}_{\alpha\beta}^{\gamma\delta}\,
{G^{(1)}}_{\gamma\delta}^{ij}=
\frac{e^{i\beta}\epsilon_{\alpha\beta}\epsilon^{ij}(z_2+S(z_1,z_2)z_1)}{(1+z_1 z_2)}
\end{equation}
and fixes the only non vanishing components of the $S$-matrix at order $g$
%This linear relation allows us to fix one of the only two non vanishing first order $S$-matrices
\begin{align}
{S^{(1)}}_{\alpha\beta}^{ij} &= e^{i\beta}
\epsilon_{\alpha\beta}\epsilon^{ij} P(z_1,z_2)= e^{i\beta}
\epsilon_{\alpha\beta}\epsilon^{ij}
\frac{(1-z_1)(1-z_2)(z_2-z_1)}
{z_1 z_2 (1-2z_1+z_1 z_2)}
\\
{S^{(1)}}_{ij}^{\alpha\beta} &= e^{-i\beta}
\epsilon^{\alpha\beta}\epsilon_{ij} Q(z_1,z_2)=e^{-i\beta}
\epsilon^{\alpha\beta}\epsilon_{ij}\frac{(1-z_1)(1-z_2)(z_2-z_1)}
{(1-2z_1+z_1 z_2)}
\end{align}
where we use again $z_j = e^{ip_j}$.

We now reinstate the boundaries of the chain and ask about the integrability of the problem by evaluating the boundary Yang-Baxter equation. Since the boundary condition is not supersymmetric and scalar and fermions got reflected with different phases it is
an interesting question whether the boundary Yang-Baxter equation is fulfilled or not.

At order $g^0$, the components of the boundary Yang-Baxter equation which are not trivially verified
correspond to the cases in which the incoming particles are one bosonic and the other fermionic. In such cases the boundary Yang-Baxter condition reads\footnote{After having used that $S(-p_2,-p_1) = S(p_1,p_2)$.}
\begin{align}
&\left(R_\phi(p_1)R_\phi(p_2)-R_\chi(p_1)R_\chi(p_2)\right)\left(S(p_1,p_2)-1\right)\left(S(-p_2,p_1)+1\right)
\\
&=\left(R_\chi(p_1)R_\phi(p_2)-R_\phi(p_1)R_\chi(p_2)\right)\left(S(p_1,p_2)+1\right)\left(S(-p_2,p_1)-1\right)
\nonumber
\end{align}
Although it looks non-trivial, for the scattering phases given in (\ref{RMatrixsu23}) this equation is satisfied, even without specifying the actual value of $\alpha$ which, as test for the integrability of our specific problem, does not seem so stringent.

However at order $g^1$, we find the boundary Yang-Baxter is not satisfied unless the parameter $\alpha$ takes some specific values. In particular, when two bosonic impurities of different flavour are in the in-going states we obtain the following condition
\begin{align}
R_\phi(p_1) \left[Q(-p_1,p_2)R_\chi(p_2)S(-p_2,-p_1)-
 R_\phi(p_2)Q(-p_2,-p_1)\right] = \nonumber
\\
R_\chi(p_1) \left[Q(p_1,p_2)R_\phi(p_2)-
R_\phi(p_2)Q(-p_2,p_1)\right]
\end{align}
which, after replacement of the scattering functions, implies that
\begin{equation}
(1-\alpha)(1-2\alpha) \frac{(1+z_1)(1-z_1^2)(1+z_2)(1-z_2^2)(z_1-z_2)(1-z_1 z_2)}{(1-2z_1 +z_1 z_2)(2-z_1-z_2)(1-\alpha-z_1)(1-\alpha-z_2)} = 0
\end{equation}
Analogous conditions are obtained for other components. Therefore, only when the boundary term parameter is either
\begin{equation}
\alpha=1\quad \mbox{or}\quad \alpha=\tfrac{1}{2}
\end{equation}
integrability can persist to this order. As we have seen from the perturbative computation, for the case of a boundary set  by an ordinary $\zeta = 0$ WL, one has precisely $\alpha=\tfrac{1}{2}$ implying non-trivially that there is integrability up to this intermediate order.

\section{Conclusions}

In this paper we have derived the 1-loop dilatation  operator for $SO(6)$ scalar composite insertions in the straight $W^{(\zeta)}$ Wilson loop, which interpolates between the ordinary WL and the supersymmetric WML.

By considering the boundary Yang-Baxter equation we obtained the main result of our paper: the corresponding 1-loop spin chain is integrable for $\zeta = \pm 1$ and $\zeta=0$, the fixed points in the renormalization group flow of the one-dimensional  defect CFT, while it is not integrable for other intermediate values.
Although this is not conclusive evidence, this result is a compelling hint that the mixing of insertions in the ordinary non-supersymmetric Wilson loop might be an integrable problem.

In order to confirm such a remarkable property one could explore the action of the dilatation operator in more general sectors and to higher loop orders.  Since the ordinary WL is non-supersymmetric, a more satisfying piece of evidence in favour of integrability would be if it persists when the inserted operators are built of fermions as well as bosons. Some steps towards this direction have also been taken in the present paper. More precisely we have derived the 1-loop dilatation  operator for $SU(2|3)$ composite insertions in the ordinary WL an verified that the boundary Yang-Baxter equations holds, even when $\lambda^{3/2}$ terms -which allow the number of fields in the composite insertions to change- are included.

If further evidence pointing towards integrability of the ordinary Wilson loop were found, one could proceed with a bootstrap program to determine the boundary reflection matrix. Moreover, it might be possible to develop a Thermodynamic Bethe Ansatz formalism to describe the corresponding cusp anomalous dimension, related to the quark/antiquark potential and the Bremsstrahlung function.

\acknowledgments

The work of M.L. is supported in part by ANPCyT (Argentina) through grant PICT-2015-1633. The work of D.C. is supported in part by grants PIP 0681, and PID {\it B\'usqueda de nueva F\'\i sica} and UNLP X850.

\appendix
\section{Boundary contributions to the $SU(2|3)$ Hamiltonian}\label{app:Integrals}

In this appendix we derive the $\alpha=1/2$ value that occurs in the boundary part of the 1-loop Hamiltonian for the $SU(2|3)$ chain. For this we have to compute the singularities of the contributions from a gluon exchange between the first/last fermion operator with the Wilson loop as in figure \ref{fig2:Ha} and \ref{fig2:Hc} and half the contribution of the self-energy diagrams of the fermion fields as in \ref{fig2:Hb} and \ref{fig2:Hd}. We begin with the former contributions. We insert the operators and the Wilson loop gluon insertion at $x_i=(\tau_i,0,0,0)$ in an orderly fashion. From Feynman rules we obtain
\begin{equation}\label{eqfig2:Ha}
\ref{fig2:Ha}=
\frac{i\lambda N \Gamma^3(1-\epsilon)}{4(4\pi^{D/2})^3}
(\sigma^{\mu})_\alpha^{\ \dot{\alpha}}
(\tilde\sigma^0)_{\dot{\alpha}}^{\ {\beta}}
(\sigma^{\nu})_\beta^{\ \dot{\beta}}
 \int_{\tau_1}^{\tau_3}d\tau_2\,
 \partial_{\mu}^{(1)}\partial_{\nu}^{(3)}
\int d^D{x_0}\frac{1}{(x_{01}^2)^{1-\epsilon}(x_{02}^2)^{1-\epsilon}(x_{03}^2)^{1-\epsilon}}
\end{equation}
Using the identity
\begin{equation}\label{SigmaIdentity}
(\sigma^{\mu})_\alpha^{\ \dot{\alpha}}
(\tilde\sigma^\rho)_{\dot{\alpha}}^{\ {\beta}}
(\sigma^{\nu})_\beta^{\ \dot{\beta}}=
\delta^{\mu\nu}(\sigma^{\rho})_\alpha^{\ \dot{\beta}}
-\delta^{\mu\rho}(\sigma^{\nu})_\alpha^{\ \dot{\beta}}
-\delta^{\rho\nu}(\sigma^{\mu})_\alpha^{\ \dot{\beta}}
+\epsilon^{\mu\nu\rho\sigma}(\sigma_{\sigma})_\alpha^{\ \dot{\beta}}
\end{equation}
we may separate the contractions with the derivatives in terms of the direction of the Wilson-line, which we call $0$, and the other directions
\begin{equation}
(\sigma^{\mu})_\alpha^{\ \dot{\alpha}}
(\tilde\sigma^0)_{\dot{\alpha}}^{\ {\beta}}
(\sigma^{\nu})_\beta^{\ \dot{\beta}}\partial_{\mu}^{(1)}\partial_{\nu}^{(3)}=
-(\sigma^{0})_\alpha^{\ \dot{\beta}}\partial_{0}^{(1)}\partial_{0}^{(3)}
-(\sigma^{i})_\alpha^{\ \dot{\beta}}
(\partial_{0}^{(1)}\partial_{i}^{(3)}+\partial_{i}^{(1)}\partial_{0}^{(3)})
+(\sigma^{0})_\alpha^{\ \dot{\beta}}
\partial_{i}^{(1)}\partial_{i}^{(3)}
\end{equation}
It is easy to see that the contribution from the mixed derivatives vanishes since it becomes the integral of an odd-function. On the other hand, it is possible to see that the contribution from the term with both derivatives in the transverse directions is finite. Thus we are left with
\begin{equation}\label{step2}
\ref{fig2:Ha}=
-\frac{i\lambda N \Gamma^3(1-\epsilon)}{4(4\pi^{D/2})^3}
(\sigma^{0})_\alpha^{\ \dot{\beta}}
\int_{\tau_1}^{\tau_3}d\tau_2\frac{\partial}{\partial\tau_1}
\frac{\partial}{\partial\tau_3}f(\tau_1,\tau_2,\tau_3)
\end{equation}
where $f(\tau_1,\tau_2,\tau_3)$ is the result of the space-time integral in (\ref{eqfig2:Ha}). We can further separate this last piece with the formula
\begin{equation}
\int_{\tau_1}^{\tau_3}d\tau_2\frac{\partial}{\partial\tau_1}
\frac{\partial}{\partial\tau_3}f(\tau_1,\tau_2,\tau_3)=
\frac{\partial}{\partial\tau_1}
\frac{\partial}{\partial\tau_3}\int_{\tau_1}^{\tau_3}d\tau_2 f(\tau_1,\tau_2,\tau_3)
+\frac{\partial}{\partial\tau_3} f(\tau_1,\tau_1,\tau_3)
-\frac{\partial}{\partial\tau_1} f(\tau_1,\tau_3,\tau_3)
\end{equation}
and we can discard the contribution of the first term on the RHS of this last formula since its easily seen to be finite. The rest of the computation becomes simpler now since the space-time integral has to be computed at coincident points
\begin{equation}
\frac{\partial}{\partial\tau_3}
\int d^{D}x_0\frac{1}{(x_{01}^2)^{2-2\epsilon}(x_{03}^2)^{1-\epsilon}}
-\frac{\partial}{\partial\tau_1}
\int d^{D}x_0\frac{1}{(x_{01}^2)^{1-\epsilon}(x_{03}^2)^{2-2\epsilon}}
=2\pi^{D/2}(-2+4\epsilon)\frac{G[2-2\epsilon,1-\epsilon]}{(\tau_3-\tau_1)^{3-4\epsilon}}
\end{equation}
where $G[a,b]$ is defined by
\begin{equation}
\int d^{D}x_0\frac{1}{(x_{01}^2)^{a}(x_{02}^2)^{b}}=
\frac{\pi^{D/2}G[a,b]}{(x_{12}^2)^{a+b-D/2}},\quad
G[a,b]=\frac{\Gamma(\tfrac{D}{2}\!-\!a)\Gamma(\tfrac{D}{2}\!-\!b)\Gamma(a\!+\!b\!-\!\tfrac{D}{2})}{\Gamma(a)\Gamma(b)\Gamma(D\!-\!a\!-\!b)}
\end{equation}
Inserting these results in (\ref{step2}) and expanding in $\epsilon$ we obtain
\begin{equation}
\ref{fig2:Ha}=T_{\alpha}^{\dot\beta}(\tau_3-\tau_1)\frac{\lambda}{16\pi^2\epsilon}
\end{equation}
where $T_{\alpha}^{\dot\beta}(\tau_3-\tau_1)$ is the tree level propagator mentioned in the main text. Thus, diagram \ref{fig2:Ha} together with its analogous \ref{fig2:Hc} contribute to the boundary Hamiltonian
\begin{equation}\label{HbdrySU23.1}
H_{\mathrm{bdry}}|_{\ref{fig2:Ha}+\ref{fig2:Hc}}=
-\frac{\lambda}{16\pi^2} (\mathcal{Q}_F^{(1)}+\mathcal{Q}_F^{(L)})
\end{equation}

Now we have to consider half the self-energy diagrams of the fermion field as in \ref{fig2:Hb} and \ref{fig2:Hd}. These contain the contribution of a loop correction to the $\chi$ propagator which can be of two types: a gluon-$\chi$ bubble or a $\psi_i$-$\phi^i$ bubble, where $\psi_i$ are the other three fermions  and $\phi^i$ are the three complex scalars of $\mathcal{N}=4$ SYM. We choose to compute this correction in momentum space and then compare it to the momentum space tree-level propagator. The contribution from the gluon correction is
\begin{equation}
-\frac{\lambda N}{2}
(\sigma^{\mu})_\alpha^{\ \dot{\alpha}}
(\tilde\sigma^\nu)_{\dot{\alpha}}^{\ {\beta}}
(\sigma^{\rho})_\beta^{\ \dot{\gamma}}
(\tilde\sigma_\nu)_{\dot{\gamma}}^{\ {\gamma}}
(\sigma^{\sigma})_\gamma^{\ \dot{\beta}}
\frac{p_\mu p_\rho}{(p^2)^2}
\int \frac{d^D{k}}{(2\pi)^D}\frac{k_\sigma}{k^2 (k-p)^2}
\end{equation}
Using identity (\ref{SigmaIdentity}), integrating in momentum space and expanding in $\epsilon$ we obtain
\begin{equation}
\frac{N\lambda G[1,1]}{2(4\pi)^{d/2}(p^2)^{1+\epsilon}}(\sigma^\mu)_{\alpha}^{\ \dot\beta}p_\mu=
-\frac{\lambda}{16\pi^2\epsilon}T_{\alpha}^{\dot\beta}(p)\label{gluonContr}
\end{equation}
where $T_{\alpha}^{\dot\beta}(p)$ is the tree-level propagator in momentum space. For the scalar triplet correction we obtain instead
\begin{align}
&-3\lambda N
(\sigma^{\mu})_\alpha^{\ \dot{\alpha}}
(\tilde\sigma^\nu)_{\dot{\alpha}}^{\ {\beta}}
(\sigma^{\rho})_\beta^{\ \dot{\beta}}
\frac{p_\mu p_\rho}{(p^2)^2}
\int \frac{d^D{k}}{(2\pi)^D}\frac{k_\nu}{k^2 (k-p)^2}\nonumber\\
&=\frac{3N\lambda G[1,1]}{2(4\pi)^{d/2}(p^2)^{1+\epsilon}}(\sigma^\mu)_{\alpha}^{\ \dot\beta}p_\mu=
-\frac{3\lambda}{16\pi^2\epsilon}T_{\alpha}^{\dot\beta}(p)\label{scalarContr}
\end{align}
Combining contributions (\ref{scalarContr}) and (\ref{gluonContr}) and taking into account that the boundary Hamiltonian only `uses' half of the self-energy corrections we obtain for diagrams \ref{fig2:Hb} and \ref{fig2:Hd}
\begin{equation}\label{HbdrySU23.2}
H_{\mathrm{bdry}}|_{\ref{fig2:Hb}+\ref{fig2:Hd}}=
\frac{\lambda}{8\pi^2} (\mathcal{Q}_F^{(1)}+\mathcal{Q}_F^{(L)})
\end{equation}
such that summing (\ref{HbdrySU23.1}) and (\ref{HbdrySU23.2}) we obtain the boundary hamiltonian of the main text (\ref{HbdrySU23}).

\end{document}